\documentclass[aps,pra,preprint,superscriptaddress]{revtex4}
\usepackage[dvips]{graphics}
\usepackage{graphicx}
\begin{document}
%
\title{Array of planar Penning traps as a nuclear magnetic 
       resonance molecule for quantum
       computation}
\author{G. Ciaramicoli}
\affiliation{Dipartimento di Fisica, Universit\`a
degli Studi di Camerino, 62032 Camerino, Italy}
\author{F. Galve}
\affiliation{Institut f\"{u}r Physik, Johannes-Gutenberg-Universit\"{a}t,
D-55099 Mainz, Germany}
\author{I. Marzoli}
\affiliation{Dipartimento di Fisica, Universit\`a
degli Studi di Camerino, 62032 Camerino, Italy}
\author{P. Tombesi}
\affiliation{Dipartimento di Fisica, Universit\`a
degli Studi di Camerino, 62032 Camerino, Italy}
\date{\today}
\begin{abstract}
An array of planar Penning traps, holding single electrons, can realize
an artificial molecule suitable for NMR-like quantum information processing.
The effective spin-spin coupling is accomplished by applying a magnetic
field gradient, combined to the Coulomb interaction acting between the
charged particles.
The system lends itself to scalability, since the same substrate can easily
accomodate an arbitrary number of traps. Moreover, the coupling
strength is tunable and under experimental control.
Our theoretical predictions take into account a realistic setting,
within the reach of current technology.
\end{abstract}
%
\pacs{}
\maketitle
\section{Introduction}
In this proposal we bring together the best of two avenues to
quantum information processing: nuclear magnetic resonance (NMR)
\cite{jones} and ion trapping \cite{leibfried1}. Both approaches
have provided the first experimental demonstrations of fundamental
quantum logic gates and quantum algorithms, though still limited
to few qubits. Actually, it is relatively simple to build a small
NMR quantum computer, useful for a proof-of-principle test of
quantum algorithms, but unable to perform any real computation,
involving hundreds of qubits. Indeed, the scalability seems to be
a rather tough question for NMR quantum computing. Another
disadvantage of NMR quantum computation is that NMR experiments
deal with a large number of molecules, building up an ensemble of
indistinguishable quantum computers. This fact brings in relevant
theoretical and practical implications: from the system
initialisation to the debate on the same quantum character of the
computation carried out with such a device. However, NMR
techniques, based on radio frequency (rf) and microwave (mw)
pulses, enable to carefully prepare, manipulate, and detect the
qubit states with relative ease. Spatial separation of the qubits
is not required, since different qubits are distinguished using
different resonance frequencies.

On the other hand, in ion trap quantum computation, qubits are stored
in isolated quantum systems, arranged to form strings of trapped particles,
spatially separated and singly addressable with optical radiation
\cite{rainerblatt,Leibfried}.
Such systems can be prepared in their motional ground state via
sophisticated
laser cooling techniques. Coherent manipulation of the qubits
requires strongly focused pulses of controlled
intensity, phase and duration at optical frequencies.
The required setup is rather involved and the experimental realization
challenging.
These technical problems have motivated other proposals by Wunderlich
\emph{et al.} to implement
a quantum computer, based on trapped ions in a linear Paul trap, 
but using long-wavelength
radiation, in the radio frequency or microwave range \cite{mintert}.
To this end, internal and external degrees of freedom of the trapped
ion are coupled by means of a magnetic field gradient.
Moreover,
in a linear Paul trap, the collective vibrational modes of
a chain of $N$ two-level ions extend this coupling to different ions.
The system is formally analogous to a collection of spins, interacting
through the so-called $J$-coupling, typical of nuclear spins in
molecules \cite{wunderlich_1,wunderlich_2,wunderlich_3}.
However, the proposal by Wunderlich \emph{et al.} presents some drawbacks,
especially for the scalability,
due to the fact that all the ions are stored in the same linear trap.
A more flexible design would be based on a string of individually
tailored microtraps, each of them trapping a single ion \cite{mchugh}.
These ideas are versatile and can be adapted to other scenarios.

Our system consists of an array of Penning traps, each of them
confining a single electron. A Penning trap makes use of static
electric and magnetic fields to trap charged particles, like ions or
electrons.
In particular, the magnetic field provides the radial confinement,
which, in a Paul trap, 
is achieved
by means of an oscillating (radio frequency) electric field.
The resulting dynamics of a particle in a Paul trap is harmonic
in all directions, whereas in a Penning trap the combination of the
electric and magnetic fields gives rise to a more complicated 
orbital motion, which is the superposition of the magnetron and
cyclotron oscillators. 
These differences should be taken into account, when describing
the effects of an additional magnetic gradient on the trapped particle
dynamics.

We choose to trap electrons instead of ions, because of their smaller
mass, which results in 
higher trapping frequency for the quantized external degrees of freedom.
Indeed, the typical resonance frequencies of the resulting electron motion
lie in the radiofrequency and microwave domain, making it possible to
employ the same technological resources and methods developed for
NMR experiments.
Moreover, differently
from a Paul trap, a Penning trap does not rely on rf
fields, a benefit in terms of stability of the trapping potential.
In turns, this translates into less decoherence affecting the
trapped particles.
Therefore, electrons trapped in vacuum seem to be a promising candidate
for quantum information processing \cite{ciaramicoli4, ciaramicoli_5},
taking  advantage of the techniques and strategies devised
both for NMR and ion trapping quantum computing.

What we have in mind is a new concept of planar Penning traps
\cite{stahl}, characterised by an open geometry. The trap
electrodes are deposited on a ceramic substrate by means of
well-established thin film or thick film technology, which allows
for a variety of different configurations as well as dimensions.
The trapping mechanism relies, as in conventional Penning traps,
on the application of a magnetic field together with an
electrostatic quadrupole potential. A single trapped electron is
confined in vacuum at an adjustable distance from the trap
surface. The same substrate can accomodate several traps in order
to form a regular one or two dimensional array of trapped
particles. Qubits are encoded in the natural two-level system provided by
the electron spin in the external strong magnetic field, similarly
to what happens for NMR spin one-half nuclei. The two possible
spin orientations $|\uparrow\rangle$ and $|\downarrow\rangle$
represent, respectively, the logic states $|1\rangle$ and
$|0\rangle$. Here, however, the two spin levels are greatly
separated in energy and thermal excitation is completely
negligible, especially at the trap cryogenic temperatures. Hence,
the system, after initialisation, remains in its ground state, corresponding
to the spin-down state. 
We have already
observed that in order to make the spin qubits distinguishable,
one has to introduce a small magnetic gradient
\cite{ciaramicoli4}. 
Actually, with a judicious choice of the magnetic gradient, that
in our configuration depends on all the spatial coordinates, we
can also build up a molecule-like system. 
In the present geometry, the magnetic field
gradient is applied across the substrate in order to differentiate
among the spin resonance frequencies at each trapping site.
Therefore, the spin qubits are distinguishable and can be
selectively frequency addressed via microwave pulses. The same
magnetic field gradient, mediated by the long range Coulomb
interaction, enables the effective spin-spin coupling between
different electrons. The resulting system may be regarded as an
artificial molecule suitable for NMR quantum computation.
Actually, we can even envisage applications to simulate other
quantum systems, like the Ising model. Our system offers obvious
advantages in terms of scalability. In addition, the spin-spin
coupling depends on external parameters, like the strength of the
magnetic field gradient, the trapping frequencies, and the trap
separation, that can be adjusted to obtain the optimal performance
of the quantum processor. 
We point out that this coupling is proportional to $1/d^3$, with $d$
being the inter-trap distance. Therefore, the spin-spin coupling strength
is relevant only for nearest-neighbor electrons, while rapidly 
decays along the electron chain.

Finally, for what concerns the
initialisation of the system, or, in other words, how to reset the
quantum register to the state $|0\rangle$, several strategies
could be followed. A possible one is to apply a transverse
oscillating magnetic field resonant with the frequency difference
between the cyclotron motion and the spin precession. When the
magnetic gradient is off the transverse field will flip the spin
up state of each particle transfering the energy to the
corresponding cyclotron motion, which will release its energy to
the environment via syncrotron radiation.

The manuscript is organized as follows.
In Sec. \ref{trap} we describe the trap design and how to
create a scalable device, putting several planar traps on the same
substrate.
The theoretical framework is developed in Sec. \ref{molecule}, where
we discuss the role of the applied magnetic field gradient to achieve
the individual frequency addressability of the qubits (Sec. \ref{gradient})
and derive an analytical expression for the effective spin-spin coupling 
(Sec. \ref{coupling}).
As an example we illustrate how to implement a set of fundamental
quantum logic gates with the resulting NMR-like molecule, made out
of trapped electrons (Sec. \ref{gates}).
The concluding remarks and the future perspectives for the
system are summarized
in Sec. \ref{conclusion}.
\section{\label{trap} The planar Penning trap}
\begin{figure}[b]
\includegraphics[scale=0.5]{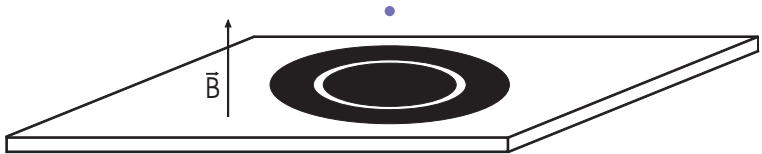}
\caption{ \label{planar_trap} Simplest configuration of a planar trap with two
electrodes (black) on an isolator substrate (white).}
\end{figure}

We have tried to design a trap which fulfills the
needs of quantum computation and at the same time ensures the same
amount of control and precision already achieved with conventional
traps. To that end we consider a planar Penning trap, which is
shown schematically in Fig.~\ref{planar_trap}. The planar trap is a Penning trap
in the sense that both electric and magnetic static fields are
used to confine three-dimensionally a charged object. A generic
Penning trap generates a quadrupolar static electric field
together with a homogeneous and static magnetic field aligned in
the direction of the symmetry axis, which we'll assume to be $z$.
The magnetic field confines a charged particle radially, whereas the
electric field provides the confinement along the $z$ axis \cite{bossBook}. 
Since
the ideal quadrupole potential depends on the square of the
coordinates, the motion in an ideal trap can be decomposed in
three independent harmonic oscillators which are commonly
denominated by cyclotron, magnetron and axial, each of them having
different characteristic frequencies.
 When the trap is not ideal, due to imperfectly homogeneous or quadrupolar
 fields, respectively, the three eigenmotions couple and can no longer be
described as independent.
However, the effects of such a  coupling can be used for a number of
applications, see e.g. \cite{pritchard,Verdu,Djekic}.

\subsection{General trap properties}

A planar trap is a novel concept of trap \cite{stahl} which allows
for easy access with radiation, because of its open geometry, and
lends itself to form a two dimensional array on the same substrate.
In addition, it has the advantage that well-known methods
to produce and miniaturize it are available.

As shown in Fig.~\ref{planar_trap}, it consists of a collection
of circular electrodes printed on an isolating substrate. The
simplest planar structure has two different electrodes to which
voltages of opposite sign are applied. The trap is lying on the
$x-y$ plane and provides an electrostatic potential minimum along
the $z$ axis at a distance $z_0$ from the substrate. A strong
homogeneous static magnetic field of the order of few Tesla is
aligned in the $z$ direction and provides radial confinement.

\begin{figure}[h]
\includegraphics[scale=0.8]{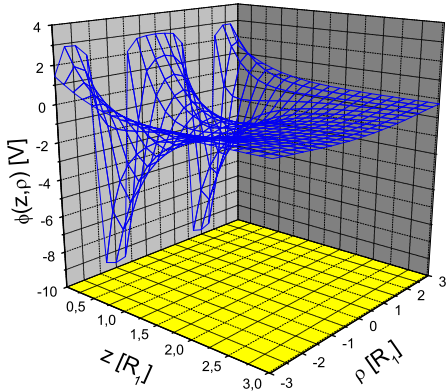}
\caption{\label{quadrupole_potential} %
Electrostatic potential with three electrodes
($V_1=3V,V_2=-10V,V_3=3V$). One can see that the local potential
minimum, in the $z$-direction around $\rho=0$, is almost
quadrupolar even with the intrinsic asymmetry of the trap.}
\end{figure}

An example of 3D potential is shown in Fig. \ref{quadrupole_potential}
 for a configuration having
three electrodes. The potential is obviously asymmetric and this
implies that one needs to fine tune the applied voltages with more
care than in a cylindrical trap. Another implication
of its asymmetry is that one can move the position of the minimum
along $z$. By varying the relative strength of the negative
electrodes with respect to positive ones, a particle which is
trapped around the minimum will vary its location under the
control of the experimenter. This feature could be useful to study
the problem of decoherence caused by a metallic plate, which
depends on the distance of the particle from the plate \cite{patch}.

The potential along the $z$ axis can, in the case of adjacent electrodes,
be obtained analitically from a Bessel-Fourier series expansion
and reads
\begin{equation}\label{zPOT}
\phi(z)=\sum_{i=1}^{N_{elec.}}
        V_i\left( \frac{z}{\sqrt{z^2+R_{i-1}^2}}
                 -\frac{z}{\sqrt{z^2+R_i^2}}
           \right),
\end{equation}
where $R_{i-1}$ is the inner radius of the $i$-th electrode
($R_0 \equiv 0$, since the first electrode is a disc). $V_i$ is
the voltage applied to electrode $i$ and $N_{elec.}$ is the number
of electrodes \cite{stahl}. The minimum of that function cannot be
evaluated analitically, but it can be shown numerically
that $z_0$, the minimum position, has a value of the order of
$R_1$, the radius of the central electrode.

For electrons lying in the region near the trap axis, $\rho \ll R_1$,
the potential is
well approximated by Eq.~(\ref{zPOT}).
Moreover, Eq.~(\ref{zPOT}) can be used
to optimise the trap geometry and enhance its harmonicity.

\subsection{Harmonicity}

In an imperfect electrostatic field one is interested in cooling
the electrons as much as possible so that they remain in the
region near the minimum, where the potential is most resembling a
quadrupolar one. A typical helium bath will thermalise electrons
to a temperature of $\sim4$~K and a dilution refrigerator at
$\sim100$~mK can drive them, on average, to the $\sim200$th energy 
level of their axial
motion, for frequencies of $\sim10$~MHz (which is the case when
$V_i\sim 0.01$~V, $R_1\sim1$~mm). The latter temperature corresponds to
an axial amplitude of $\sim20\;\mu$m and radial amplitude of the
same order of magnitude. The
amplitude of motion is thus much smaller than the characteristic
trap size $R_1$. Further cooling by pulses can be used in order
to reduce the width of the axial oscillation.

Detection of the electron axial motion can be performed by
pickup of the induced image current in the central trap electrode,
via a tuned resonance circuit of high quality factor $Q$. A
Fourier transform analysis of the induced current shows a peak at
the electron oscillatory frequency, whose width $\Delta \omega_z$
is given by the inverse time cooling constant due to the
resistivity $R$ of the detection electronics \cite{diederich}.
With values for the tank circuit such as $Q=300$, $C=7.5$~pF and
for $\omega_z/2\pi =  10$~MHz we have $\Delta
\omega_z / 2\pi \sim 1$~kHz.
Inside that peak, a dip is found, due to the electron which resonantly absorbs 
energy from the thermal noise in the circuit. 
Such a dip has a frequency width given by the axial anharmonicity. 
From numerical estimates we expect such a frequency width, for electrons 
which are thermalised to an environmental temperature of about 100~mK, 
to be, at least, two orders of magnitude narrower than that of the 
tank circuit \cite{stahl}. 

Moreover, to prevent detrimental effects on the computation due 
to electronic noise
fed into the system, it is possible to decouple the axial motion of the 
electron
from the detection circuit during gate operations. 
This is achieved by detuning either the external circuit or the
axial oscillator.  Only at the end of the computation, the axial 
frequency is brought into resonance with the detection circuit in order
to perform the final read out of the qubit states.

A superconducting solenoid provides the magnetic field along the $z$
axis. A field strength of the order of few Tesla gives a cyclotron
frequency of approximately $100$~GHz which ensures that electrons will
radiate their cyclotron energy, via synchrotron emission,
in the time scale of $0.1$~s (see
Ref.~\cite{jackson}). Experimental observation of a trapped electron
cooled down to its cyclotron ground state via radiation and in
equilibrium with its cryogenic environment has been reported by Gabrielse
and coworkers \cite{gabrielse} under conditions similar to those
mentioned here.

Furthermore there is a way of
coupling the spin motion to the axial one in order to know the
spin state from a shift in the axial frequency. Such a
coupling is realized when a quadratic magnetic gradient
is applied. The method is well known and has been
experimentally demonstrated for detection of the electron spin
state in a hydrogen-like ion \cite{bossBook}. In next sections we
will show that a gradient, which is linear in the coordinates, can
be used to provide an effective spin-spin coupling between
different electrons. Therefore, if one desires to couple electrons
and at the same time observe their spin state through measuring
the axial frequency, both linear and quadratic gradients are
needed. It is not a problem, in general, to tailor the required magnetic
field. A ring made of magnetic material, such as
Ni, has been already used for $g$-factor experiments. Such a ring
provides a quadratic gradient, when seen from its symmetry centre.
If the ring is displaced with respect to the trap axis, linear
components will be seen by the trapped electrons in addition to
quadratic ones. Other configurations are possible with the use of
additional coils.

With a typical nickel ring, the magnetic gradient is such that the 
electron axial resonance frequency suffers a shift of approximately 10~Hz 
depending on the spin state.
Considering the parameters for the tank circuit given above, 
a dip shift of 10~Hz inside a broad peak of 1~kHz is easily detectable, 
furthermore when the absorption linewidth is 10~Hz 
-- which is our case --.

\begin{figure}[h]
 \includegraphics[scale=0.4]{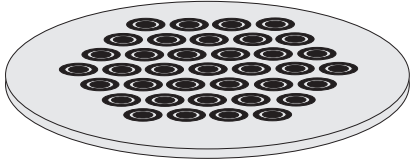}
\caption{\label{array} Two dimensional array of planar traps consisting of
printed electrodes on a base substrate made of isolating material.
Each trap is supposed to be loaded with an electron cloud which is
afterwards reduced to a single electron via a sequence of pulses.
}
\end{figure}

\subsection{Array of planar traps}

The planar structure of the novel trap
strongly
suggests a 2D array of such traps, as schematically illustrated in
Fig. \ref{array}. Thin-film technology can be
used to place gold electrodes on an isolating surface with
resolution much below the mm scale.
A large number of traps can be embedded in a common
isolating substrate and controlled electronically from the rear
side.
Ideally each trap is filled with a single electron, which interacts
with neighboring particles via the
Coulomb force. This provides a natural multiparticle
scenario similar to the case of ions in a linear Paul trap,
with the advantage of controlling each trap parameter independently
(inter-particle distance, coordination number, electron resonance
frequencies, \ldots).
With the planar trap
this is implemented in a very straightforward way.

The substrate shown in Fig. \ref{array} has the size of a coin and
each trap a dimension of order 1 mm. This size gives an axial
frequency up to 500 MHz for applied voltages of a few Volts.
The space between traps has to be filled with a common
grounded electrode, so as to isolate the potentials of each trap
from the others and also to prevent charging up of the isolating
substrate.
%
%
Thin-layer techniques allows to produce electrodes which are almost
monocristalline, and so to reduce significantly the decoherence effect
due to patch fields. On the other hand, quantitative measurements of
decoherence caused by proximity to the electrodes could be
performed by moving the electron
along the $z$ axis (see \cite{stahl}).

\begin{figure}[h]
 \includegraphics[scale=0.4]{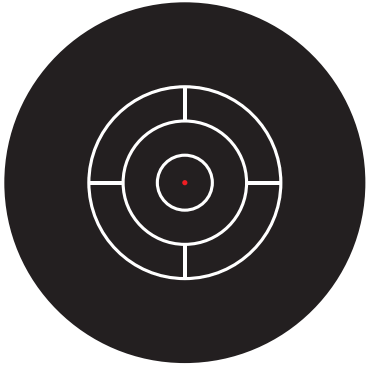}
\caption{\label{test_trap} Test trap with total diameter of 35 mm. Black regions
denote electrodes and white regions denote isolating surface. As
said before the isolating surface has been mostly covered by
electrodes to avoid charging up. This configuration has three
active electrodes plus an external grounded one. The red circle in
the center is a hole through which electrons are loaded from the
rear side.}
\end{figure}

Currently, we are testing a prototype of planar trap.
The idea is to have only one trap, with a radius of 2 cm,
printed with thick-layer technology over a Al$_2$O$_3$
substrate. A magnetic field of 100 G is foreseen. This
preliminary trap will try to demonstrate confinement
and also the possibility to excite each degree of
freedom of the electrons. Figure \ref{test_trap}
shows the design of such a test
trap. It can be seen that there is a split electrode, which will
provide quadrupolar excitation.
In a further step,
we plan to build a
miniaturised array of several traps (three for example) in a
cryogenic environment of about 100 mK with a magnetic field of 7 Tesla. The
traps will have a diameter of around 0.5 mm. The biggest challenges
will be those of coherent and accurate control over each electron
plus a sufficient supression of all sources of decoherence, both
haunting every experiment in quantum computation.
\section{\label{molecule} Building an artificial molecule}
\subsection{\label{gradient} The magnetic gradient}
\begin{figure}
 \includegraphics{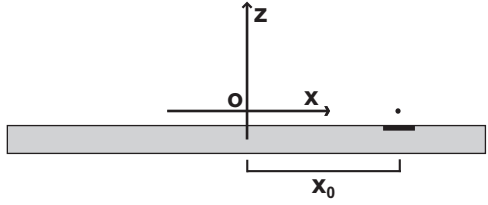}%
 \caption{\label{plan1}Schematic draw of a planar trap at the distance $x_0$
        from
        the center of the substrate.}
\end{figure}

Let us consider a planar trap at a distance $x_0$ from the center
of the substrate along the $x$-direction. We choose the origin of
our reference frame at the substrate center and the $z$-axis
orthogonal to the substrate plane (see Fig. \ref{plan1}).
Now we suppose to add the inhomogeneous magnetic field
\begin{equation} \label{magrad}
{\bf {B_1}} = b \left(z {\bf{\hat{k}}} -\frac{x}{2}
{\bf{\hat{i}}}-\frac{y}{2} {\bf{\hat{j}}}\right).
\end{equation}
This field produces a linear magnetic gradient and has rotational
symmetry with respect to the $z$-axis.\\
The total magnetic field acting on the trapped particle is $\bf
{B}=\bf {B_0}+\bf {B_1}$ where $\bf {B_0}$ is the uniform
confining field directed along the $z$-axis.
Hence, the total field $\bf {B}$ acting on the trap center is the
sum of the uniform field $B_0$ along the $z$-direction and the
magnetic gradient $-bx_0/2$ along the $x$-direction. Consequently
$\bf {B}$, at the trap center, has modulus $B_c\equiv\sqrt{B_0^2 +
(b^2x_0^2)/4}$ and direction forming an angle with the $z$-axis.
Note that $B_c$ depends on the distance $|x_0|$ of the trap from
the substrate center.
Now we rotate the reference frame around the $y$-axis so that the
new $z$-axis corresponds to the direction of the total magnetic
field $\bf {B}$ at the trap center. With this rotation, if
$b|x_0|/2 B_0 \ll 1$, the total field with respect to the new
coordinates can be written, in good approximation, as
\begin{equation} \label{magfield}
{\bf {B}} \simeq \left[(B_c + bz){\bf{\hat{k}}}
-\frac{b(x-x_0)}{2}
{\bf{\hat{i}}}-\frac{by}{2}{\bf{\hat{j}}}\right].
\end{equation}
The vector potential of the total magnetic field $\bf {B}$ applied
to the electron is
\begin{equation} \label{potvec}
{\bf {A}} \simeq
\frac{1}{2}(B_c+bz)[(x-x_0){\bf{\hat{j}}}-y{\bf{\hat{i}}}].
\end{equation}
We suppose to work with a uniform field $B_0$ of few Tesla, a
magnetic gradient $b$ of about 50~T/m and a trap substrate with
length of the order of $10^{-2}$m. Consequently we have $b|x_0|/2
B_0 < 10^{-1}$.

Let us write the Hamiltonian of the trapped electron. By taking
into account the trapping potential we have
\begin{equation}\label{Hele}
  H = \frac{(\mathbf{p} -e \mathbf{A})^2}{2m_e} +eV -
  \frac{ge\hbar}{4m_e} \mbox{ \boldmath $\sigma$ } \cdot \mathbf{B} ,
\end{equation}
where $V$, $m_e$, $e$, $g$, and $\sigma_i$ are, respectively, the
trapping potential, electron mass, electron charge, giromagnetic
factor, and Pauli matrices. In the limit $b|x_0|/2B_0 \ll 1$ we
can neglect the changes in the quadrupole potential form due to
the rotation of the reference frame around the $z$-axis 
and write
\begin{equation}\label{hj}
  V \simeq \frac{V_0}{\ell^2} \left[ z^2-\frac{(x-x_0)^2+y^2}{2}\right].
\end{equation}
We can define the axial frequency
$\omega_z\equiv\sqrt{2e V_0 / (m_e \ell^2)}$, 
in terms of the applied potential difference $V_0$ and of the 
characteristic trap size $\ell$,
and recast the $z$-part of the
spatial Hamiltonian of the electron as
\begin{equation}\label{Hzele1}
  H_z \simeq \frac{p_z^2}{2m_e}+\frac{1}{2}m_e \omega_z^2 z^2.
\end{equation}
The presence in the Penning trap of the magnetic gradient along
the $z$-direction makes the cyclotron frequency depend on the
particle $z$-position. 
Indeed, we define the cyclotron frequency as
\begin{equation}\label{wc}
\omega_c(z) \equiv \frac{|e|(B_c+bz)}{m_e}
\end{equation}
and introduce respectively the cyclotron and magnetron ladder
operators
\begin{eqnarray}
\label{ac} a_{c} &=&\frac{1}{2} \left\{ \sqrt{\frac{m
\tilde{\omega}_{c}}{2
                            \hbar}} \, [(x-x_0)-iy]+ \sqrt{\frac{2}{\hbar m
                            \tilde{\omega}_{c}}} \, (p_{y}+ip_{x})
                     \right\} , \\ \label{am}
a_{m} &=& \frac{1}{2} \left\{ \sqrt{\frac{m \tilde{\omega}_{c}}{2
                             \hbar}} \, [(x-x_0)+iy]- \sqrt{\frac{2}{\hbar m
                             \tilde{\omega}_{c}}} \, (p_{y}-ip_{x})
                      \right\} ,
\end{eqnarray}
obeying the commutation relation $[a_{i},a_{j}^{\dagger}]
=\delta_{i,j}$, with $i,j=c,m$.
The frequency $\tilde{\omega}_c$ is defined as
\begin{equation}\label{wct}
\tilde{\omega}_c(z) \equiv \sqrt{\omega_c^2 -2\omega_z^2} ,
\end{equation}
so it depends on the $z$-coordinate too.

The part of the spatial Hamiltonian of the electron involving
$x$-$y$ coordinates is
\begin{eqnarray}\label{Hxyele}
  H_{xy} &\simeq& \frac{1}{2 m_e} \left[p_x+\frac{e(B_c+bz)}{2}y\right]^2
  + \frac{1}{2 m_e} \left[p_y-\frac{e(B_c+bz)}{2}(x-x_0)\right]^2
  \nonumber \\
  &-&\frac{1}{4}m_e\omega_z^2[(x-x_0)^2+y^2] .
\end{eqnarray}
By using the ladder operators, Eqs. (\ref{ac}) and (\ref{am}), we can write
\begin{equation}\label{Hxy}
H_{xy} = -\hbar\omega_m \left(a_m^\dagger a_m + \frac{1}{2}
\right)
    + \hbar\omega'_c \left(a_c^\dagger a_c + \frac{1}{2} \right) ,
\end{equation}
where we have introduced, respectively, the magnetron and cyclotron
frequencies
\begin{eqnarray}\label{wm}
\omega_m(z) \equiv \frac{(\omega_c - \tilde{\omega}_c)}{2}, \\
\label{wpc} \omega'_c(z) \equiv \frac{(\omega_c +
\tilde{\omega}_c)}{2},
\end{eqnarray}
both depending, by means of relations (\ref{wc}) and (\ref{wct}), on the
$z$-coordinate.
Notice that we indicate the explicit dependence of the frequencies
$\omega_c$, $\tilde{\omega}_c$, $\omega'_c$, and $\omega_m$ on the
coordinate $z$ only in the definition formula.

The uniform magnetic field of few Tesla, we suppose to
apply, gives a cyclotron frequency at the trap center of the order
of $100$~GHz. This choice permits to have, after thermalization
with a trap environment at about $100$~mK, the cyclotron motion in
the lowest energy state \cite{gabrielse}. Furthermore we choose the
axial frequency in the range of MHz, giving a magnetron frequency
at the trap center of the order of hundreds of Hz.

The spatial Hamiltonian, Eq.~(\ref{Hxy}), is formally equivalent to
the corresponding part of the Hamiltonian describing an electron in
a conventional Penning trap, without any magnetic gradient \cite{brown}.
However, we point out that,  with respect to the case of a trap without 
magnetic gradient, there is a dependence of the magnetron and cyclotron
frequencies on the $z$-coordinate.
Hence, we have a coupling
between the axial motion and the cyclotron/magnetron motions.
Furthermore, as we shall see, the magnetic gradient introduces an
interaction between the spatial motion and the spin motion of the
electron.
This coupling between the external and internal degrees of freedom
becomes evident by considering the part of the electron Hamiltonian
involving the spin motion
\begin{equation}
  H_s \simeq \frac{ge\hbar}{4m_e} \mbox{\boldmath $\sigma$ } \cdot
  \mathbf{B}=\frac{g\hbar}{4}\omega_c \sigma_z-\frac{g\hbar|e|b}{8
  m_e}(\sigma_x x+\sigma_y y).
\end{equation}
We generally suppose, in our system, the \emph{axialization} of the
electron motion. This condition has been experimentally obtained
with ions confined in a Penning trap \cite{powell}. The motion of
the electron is \emph{axialized} when its amplitude in the $x$-$y$
direction is much smaller than that in the $z$-direction.
The axialization permits to neglect, in the above Hamiltonian, the
term proportional to $\sigma_x x+\sigma_y y$. Indeed, with the
typical values chosen for $b$ and $B_0$, the spin state
transition probability due to this term is negligible in the
axialization condition. This can be proved by applying
time-dependent perturbation theory.

Hence, the global Hamiltonian (\ref{Hele}) of the electron can be
rewritten as
\begin{eqnarray}\label{Hglo}
H &\simeq&-\hbar\omega_m \left(a_m^\dagger a_m + \frac{1}{2}
\right)
    + \hbar\omega'_c \left(a_c^\dagger a_c + \frac{1}{2} \right)
+ \frac{p_z^2}{2m_e}+\frac{1}{2}m_e \omega_z^2 z^2+\frac{g\hbar
}{4}\omega_{c} \sigma_z ,
\end{eqnarray}
where the frequencies $\omega_m$, $\omega'_c$, $\omega_c$ and
$\tilde{\omega}_c$, as defined in Eqs.
(\ref{wc}), (\ref{wct}), (\ref{wm}), and (\ref{wpc}), depend on the
$z$-coordinate.

We suppose that, throughout the electron motion, the conditions
$\omega_z\ll\omega_c$ and $b|z|/2B_0 \ll 1$ are verified. Hence, by
posing $\omega'_c \simeq \omega_c$,
$\omega_m\simeq\omega_z^2/2\omega_c$ and expanding the cyclotron
and magnetron frequencies in terms of the coordinate $z$, we write
the above Hamiltonian as
\begin{eqnarray}\label{Hglo2}
H &\simeq& -\hbar\omega_{m0} a_m^\dagger a_m
    + \hbar\omega_{c0} a_c^\dagger a_c
    + \hbar\omega_z a_z^\dagger a_z+\frac{\hbar}{2}\omega_{s0} \sigma_z
    \nonumber \\
   &+& \hbar \omega_z \varepsilon
       (a_z+a_z^{\dagger})\left(\frac{\omega_{m0}}{\omega_{c0}}
        a_m^\dagger a_m
        + a_c^\dagger a_c + \frac{g}{4} \sigma_z\right) ,
\end{eqnarray}
where we have introduced the ladder operator
\begin{equation}
a_z \equiv \sqrt{\frac{m_e \omega_z}{2 \hbar} } z 
       + i \sqrt{\frac{1}{2 \hbar m_e \omega_z} } p_z,
\end{equation}
the parameter
\begin{equation}\label{eps}
\varepsilon \equiv \frac{|e|b}{m_e \omega_z} \sqrt{\frac{\hbar}{2
m_e \omega_z}},
\end{equation}
and the frequencies $\omega_{c0}\equiv(|e|B_c)/m_e$,
$\omega_{m0}\equiv \omega_z^2/ 2 \omega_{c0}$,
$\omega_{s0}\equiv(g\omega_{c0})/2$.

In Hamiltonian (\ref{Hglo2}) we have also redefined the
equilibrium position of the electron in the $z$-direction, as a
consequence of the shifts due to the zero point energy of the
cyclotron and magnetron oscillators.

The parameter $\varepsilon$ gives, substantially, the ratio
between the change in the electron spin energy due to the magnetic
gradient and the trapping energy in the $z$-direction. 
Indeed we
can write 
\begin{equation}
\label{epsilon}
          \varepsilon \simeq 
             \frac{ \hbar \partial_z \omega_s \Delta z}{\hbar \omega_z},
\end{equation}
where
\begin{equation}
       \partial_z\omega_s \equiv \frac{g|e|b}{2m_e}
\end{equation}
 and 
\begin{equation}
\label{delta_z}
      \Delta z \equiv \sqrt{ \frac{\hbar}{2m_e\omega_z} }
\end{equation}
is the amplitude of the electron
axial motion in the ground state. Hence $\partial_z \omega_s
\Delta z$ is roughly the spin frequency variation due to the
magnetic gradient when the electron axial motion is in the ground
state. 

Note that the frequencies $\omega_{c0}$, $\omega_{m0}$ and
$\omega_{s0}$ depend on the position of the planar trap on the
substrate. Indeed they depend on the distance $x_0$ of the trap
center from the symmetry axis of the magnetic gradient. For
instance, when the magnetic gradient $b\simeq 50$~T/m, two electrons
in neighboring traps, separated by a distance of the order of
$10^{-3}$~m, are characterized by spin resonance frequencies that
differ from each other of few MHz. This value, though small in
comparison with the typical spin frequency $\omega_{s0} / 2\pi \sim
100$~GHz of a single electron, is enough to individually
address the spin qubits via microwave radiation. Moreover, the
same substrate can accomodate tens of qubits, with frequencies
spread over a range of hundred MHz.

If the electron motion is axialized and the cyclotron oscillator
remains always in the ground state, we can neglect, in Hamiltonian
(\ref{Hglo2}), the term proportional to $(a_z
+a_z^\dagger)[(\omega_{m0}/\omega_{c0}) a_m^\dagger a_m
+a_c^\dagger a_c]$.
Hence, under these conditions, we
can write the Hamiltonian of the system as
\begin{eqnarray}\label{Hglo3}
H &\simeq& -\hbar\omega_{m0} a_m^\dagger a_m
    + \hbar\omega_{c0} a_c^\dagger a_c
+ \hbar\omega_z a_z^\dagger a_z+\frac{\hbar}{2}\omega_{s0}
\sigma_z + \frac{g}{4} \hbar \omega_z \varepsilon
(a_z+a_z^{\dagger}) \sigma_z  ,
\end{eqnarray}
showing the coupling between the axial and the spin degrees of
freedom.

\subsection{\label{coupling} Effective spin-spin coupling}
We now move to a linear array of $N$ electrons confined in planar
Penning traps along the $x$-axis. If we add a linear magnetic
gradient, the Hamiltonian of the system can be written as
\begin{equation}\label{Harray}
  H = \sum_{i=1}^N \left[\frac{(\mathbf{p_i} -e \mathbf{A_i})^2}{2m_e} +eV_i
-
  \frac{ge\hbar}{4m_e} {\bf {\sigma_i}} \cdot
  \mathbf{B_i}\right]+
\sum_{i>j}^{N}\frac{e^2}{4 \pi
\epsilon_0|\mathbf{r_i}-\mathbf{r_j}|},
\end{equation}
where the subscript $i$ refers to the $i$-th electron of the
array.

Hence the magnetic field $\mathbf{B_i}$, with vector potential 
$\mathbf{A_i}$, acts on the electron trapped in the $i$-th site.
This field is the sum of the uniform confining field orthogonal to
the planar trap substrate and the inhomogeneous field producing a
linear magnetic gradient [see derivation of Eq. (\ref{magfield})
in the previous subsection]
\begin{equation} \label{magfield2}
{\bf {B_i}} \simeq \left[(B_{c,i} + bz_i){\bf{\hat{k}}}
-\frac{b(x_i-x_{i,0})}{2}
{\bf{\hat{i}}}-\frac{by_i}{2}{\bf{\hat{j}}}\right].
\end{equation}
In the above equation $B_{c,i} \bf{\hat{k}}$ is the total field
$\mathbf{B_i}$ at the center of the $i$-th trap and $x_{i,0}$ is
the $x$-coordinate of the center of the $i$-th trap. The origin of
our reference frame is at the center of the trap substrate and the
$z$-axis is the symmetry axis of the magnetic gradient. We have
$B_{c,i}=\sqrt{B_0^2 + (b^2 x_{i,0}^2)/4}$ where $|x_{i,0}|$ is
the distance between the symmetry axis of the magnetic gradient
and the center of the $i$-th site. In writing Eq.
(\ref{magfield2}) we assumed that $b|x_{0,i}|/2B_0 \ll 1$ for any
$i =1,2,\ldots,N$.

Similarly to the single electron case, we also have
\begin{equation} \label{potveci}
{\bf {A}_i} \simeq
\frac{1}{2}(B_{c,i}+bz_i)[(x_i-x_{i,0}){\bf{\hat{j}}}-y_i{\bf{\hat{i}}}].
\end{equation}
and
\begin{equation}
  V_i \simeq \frac{V_0}{\ell^2}
             \left[ z_i^2-\frac{(x_i-x_{i,0})^2+y_i^2}{2}\right].
\end{equation}
Now, as already done for a single electron in a planar trap in the
presence of a magnetic gradient, we define the frequencies
$\omega_{c,i}(z_i) \equiv |e|(B_{c,i}+bz_i)/m_e$ and
$\tilde{\omega}_{c,i}(z_i) \equiv \sqrt{\omega_{c,i}^2
-2\omega_z^2}$ and introduce the operators
\begin{eqnarray}
\label{acj} a_{c,j} &=&\frac{1}{2} \left\{ \sqrt{\frac{m_e
\tilde{\omega}_{c,j}}{2
                            \hbar}} \, [(x_j-x_{j,0})-iy_j]+
\sqrt{\frac{2}{\hbar
                            m_e
                            \tilde{\omega}_{c,j}}} \, (p_{y,j}+ip_{x,j})
                     \right\} , \\ \label{amj}
a_{m,j} &=& \frac{1}{2} \left\{ \sqrt{\frac{m_e
\tilde{\omega}_{c,j}}{2
                             \hbar}} \, [(x_j-x_{j,0})+iy_j]-
\sqrt{\frac{2}{\hbar
                             m_e
                             \tilde{\omega}_{c,j}}} \, (p_{y,j}-ip_{x,j})
                      \right\} ,\\ \label{azj}
a_{z,j} &=& \sqrt{\frac{m_e \omega_z}{2 \hbar}} z_j + i
\sqrt{\frac{1}{2 \hbar m_e \omega_z}} p_{z,j}
\end{eqnarray}
obeying the commutation relation $[a_{i},a_{j}^{\dagger}]
=\delta_{i,j}$, with $i,j=c,m,z$.

If we indicate with $H^{NC}$ the part of Hamiltonian
(\ref{Harray}) not including the Coulomb interaction, we can write
\begin{eqnarray}\label{HNC}
H^{NC} &\simeq& \sum_{i=1}^{N}\left( -\hbar\omega_{m0,i}
a_{m,i}^\dagger a_{m,i}
    + \hbar\omega_{c0,i} a_{c,i}^\dagger a_{c,i}
+ \hbar\omega_z a_{z,i}^\dagger a_{z,i}+\frac{\hbar}{2}\omega_{s0,i}
\sigma_{z,i}\right) \nonumber \\
&+& \sum_{i=1}^{N} \frac{g}{4} \hbar \omega_z \varepsilon
(a_{z,i}+a_{z,i}^{\dagger}) \sigma_{z,i} ,
\end{eqnarray}
where we have introduced the frequencies
$\omega_{c0,i}\equiv(|e|B_{c,i})/m_e$, $\omega_{m0,i}\equiv
\omega_z^2/ 2 \omega_{c0,i}$,
$\omega_{s0,i}\equiv(g\omega_{c0,i})/2$ and assumed that,
throughout the electron motion, the conditions
$\omega_z\ll\omega_{c,i}$ and $b|z_i|/2B_0 \ll 1$ are verified for
any $i =1,2,\ldots,N$. The parameter $\varepsilon$ has been defined 
in Eq. (\ref{eps}).
In writing Hamiltonian (\ref{HNC}) we have also supposed that all
the electron motions are axialized and that the cyclotron
oscillator remains always in the ground state.

Let us consider the part of Hamiltonian (\ref{Harray}) involving the
Coulomb interaction between the two electrons at the sites $i$ and
$j$. By indicating it with $H^{C}_{i,j}$ we have
\begin{equation}\label{HC}
  H^{C}_{i,j} = \frac{e^2}{4 \pi \epsilon_0 \sqrt{(x_i-x_j)^2 +
            (y_i-y_j)^2+(z_i-z_j)^2}},
\end{equation}
which we can recast as
\begin{eqnarray}\label{HC2}
   H^{C}_{i,j} = \frac{e^2}{4 \pi \epsilon_0 d_{i,j}} \left[1+\frac{2(\Delta
x_i-\Delta x_j )}{d_{i,j}}+
  \frac{(\Delta x_i-\Delta x_j
)^2}{d_{i,j}^2}+\frac{(y_i-y_j)^2}{d_{i,j}^2}+
  \frac{(z_i-z_j)^2}{d_{i,j}^2} \right]^{-\frac{1}{2}} ,
\end{eqnarray}
where $\Delta x_i\equiv x_i -x_{i,0}$ and
$d_{i,j}=|x_{i,0}-x_{j,0}|$.

If the oscillation amplitude of the two electrons is much smaller
than the average separation $d_{i,j}$ between them, we can expand
the interaction Hamiltonian, Eq.~(\ref{HC2}), in a power series
and retain the terms up to the second order
\begin{equation}\label{HC3}
  H^{C}_{i,j} \simeq -\frac{e^2}{4 \pi \epsilon_0 d_{i,j}^2} (\Delta
x_i-\Delta x_j) +
  \frac{e^2}{8 \pi \epsilon_0 d_{i,j}^3} \left[ 2(\Delta x_i-\Delta x_j)^2 -
(y_i-y_j)^2
                               - (z_i-z_j)^2 \right].
\end{equation}
Furthermore, if we suppose that the electron motions are
axialized, we have
\begin{equation}\label{HC4}
  H^{C}_{i,j} \simeq -\frac{e^2}{4 \pi \epsilon_0 d_{i,j}^2} (\Delta
x_i-\Delta x_j)
  -
  \frac{e^2}{8 \pi \epsilon_0 d_{i,j}^3} (z_i-z_j)^2.
\end{equation}
The first term in Hamiltonian (\ref{HC4}) gives a displacement of
the equilibrium position of the electrons along the $x$-axis.
This small shift is of order of $e^2/(4 \pi \epsilon_0 d_{i,j}^2
m_e \omega_{c0,i}^2)$ and is a consequence of the Coulomb
repulsion. Hence we can effectively remove the first term in
Hamiltonian (\ref{HC4}) by redefining the centers of the two
traps. We also recall that, actually, this effect is more pronounced
for electrons placed at the extremities
of the array. Indeed, as a consequence of the symmetry of the
system, particles trapped near the center of the array undergo much
smaller shifts.

The second term in Hamiltonian (\ref{HC4}), involving the
$z$-coordinate of the electrons, represents an effective dipole-dipole 
interaction between the $i$-th and $j$-th electrons. 
By developing the square we have
terms proportional to $z_i^2$ and $z_j^2$. They produce small
shifts on the axial frequencies of the two electrons. Therefore we can
take into account these terms by appropriately redefining the
axial frequencies of the two electrons. The remaining term,
proportional to $z_i z_j$ represents the Coulomb coupling between
the axial motion of the two electrons. After having appropriately
redefined the trap centers and frequencies we can write
\begin{equation}\label{HC5}
  H^{C}_{i,j} \simeq \frac{e^2}{4 \pi \epsilon_0 d_{i,j}^3} z_i
  z_j = \hbar \xi_{i,j}
  (a_{z,i}+a^{\dagger}_{z,i})(a_{z,j}+a^{\dagger}_{z,j}) ,
\end{equation}
where
\begin{equation}
\hbar \xi_{i,j} \equiv \frac{e^2}{4 \pi \epsilon_0 d_{i,j}} \left(
\frac{\Delta z}{d_{i,j}}\right)^2,
\end{equation}
with $\Delta z$ being the ground state amplitude of the axial 
oscillator, introduced in Eq.~(\ref{delta_z}).

Now we perform, on the global Hamiltonian of the system
$H=H^{NC}+\sum_{i>j}^N H^{C}_{i,j}$, the unitary transformation
$H'=e^{S}He^{-S}$ \cite{mintert} with
\begin{equation}     \label{S}
  S = \sum_{i=1}^N \varepsilon  \frac{g}{4}
      (a_{z,i}^\dagger-a_{z,i}) \sigma_{z,i}.
\end{equation}
The transformed axial operators are
\begin{equation}
a_{z,i} \rightarrow  a_{z,i} - \varepsilon  \frac{g}{4} \sigma_{z,i}.
\end{equation}
Let us consider the transformed part of the Hamiltonian
(\ref{Harray}) not including the Coulomb terms. It can be written,
after dropping constant terms, as
\begin{eqnarray}\label{HNCp}
H'^{NC} &\simeq& \sum_{i=1}^{N}\left( -\hbar\omega_{m0,i}
a_{m,i}^\dagger a_{m,i}
    + \hbar\omega_{c0,i} a_{c,i}^\dagger a_{c,i}
+ \hbar\omega_z a_{z,i}^\dagger a_{z,i}+
\frac{\hbar}{2}\omega_{s0,i} \sigma_{z,i}\right)
\end{eqnarray}
Note that, with the unitary transformation, we have formally
removed the interaction between the axial and spin motions.
Differently the Coulomb terms in Hamiltonian (\ref{Harray})
involving the electrons $i$ and $j$ transform as
\begin{eqnarray}\label{HCp}
  H'^{C}_{i,j} &=& \hbar \xi_{i,j}
  \left( a_{z,i}+a^{\dagger}_{z,i}- \varepsilon
         \frac{g}{2} \sigma_{z,i}
  \right) 
  \left( a_{z,j}+a^{\dagger}_{z,j}- \varepsilon
         \frac{g}{2} \sigma_{z,j}
  \right).
\end{eqnarray}
From the above Hamiltonian we see that the unitary transformation
produces terms of the form $\hbar (g^2/4) \varepsilon^2 \xi_{i,j}
\sigma_{z,i} \sigma_{z,j}$, which represent an effective
coupling between the spin motion of different particles.

By applying another unitary transformation, similar to Eq.~(\ref{S}),
also the extra terms in  Hamiltonian (\ref{HCp}),
proportional to $\sigma_{z,i}(a_{z,j}+a^{\dagger}_{z,j})$, 
result in additional spin-spin coupling terms. 
As a consequence we
would have corrections to the coupling strength of the spin-spin interaction
the order of $\hbar \varepsilon^2 \xi_{i,j}^3/\omega_z^2$.
However, we assume $\xi_{i,j} \ll \omega_z$, so that the effect due to the
terms proportional to $\sigma_{z,i}(a_{z,j}+a^{\dagger}_{z,j})$
is negligible.

Since in our scheme quantum information is encoded only in the
spin motion of the particles, we do not consider the spatial part
of the transformed Hamiltonian, but just the spin-part, given by
\begin{eqnarray}\label{Heff}
H'_s &\simeq& \sum_{i=1}^{N}\frac{\hbar}{2}\omega_{s0,i}
\sigma_{z,i} + \sum_{i>j}^N \frac{\hbar}{2} \pi J_{i,j}
\sigma_{z,i} \sigma_{z,j}.
\end{eqnarray}
Note that the above Hamiltonian is analogous to the nuclear spin
Hamiltonian of the molecules used to perform NMR quantum
computation \cite{jones}.
Consequently, with our system, we can implement universal quantum
processing by using the same techniques developed in NMR experiments
(see next section).

In Hamiltonian (\ref{Heff}) the spin frequencies are
\begin{equation}
\omega_{s0,i}\simeq\frac{geB_0}{2m_e}\left(1+\frac{b^2x_{i,o}^2}{8B_0^2}\right)
\end{equation}
and the coupling constants, defined as in NMR experiments, are
\begin{equation}\label{J}
J_{i,j}\equiv \frac{g^2}{2\pi} \xi_{i,j}\varepsilon^2 =
\frac{g^2}{2\pi\hbar} \frac {e^2}{4\pi\epsilon_0 d_{i,j}} \left(\frac
{\Delta z}{d_{i,j}}\right)^2 \left(\frac{\hbar \partial_z \omega_s
\Delta z}{\hbar \omega_z}\right)^2 \propto \frac{b^2}{\omega_z^4
d_{i,j}^3}.
\end{equation}
Hence, by changing specific system parameters, fully under the
control of the experimenter, we can adjust both the spin frequencies
and the coupling constants.
Indeed, according to the above relations, the spin frequencies of
the particles and, consequently, their detunings depend on the
uniform magnetic field intensity, the magnetic gradient strength and 
the distance of
the electrons from the substrate center.
Similarly, the coupling constants can be modified by
changing  the gradient strength, the
axial frequency and the distance between the particles.
We also note that the coupling constants are proportional to
$1/d_{i,j}^3$. Hence, if the particles are equally spaced in the
array, we obtain a uniform coupling strength for nearest-neighbor
electrons, while the spin-spin interaction 
decreases rapidly with the increase of their
distances. 
For instance, the coupling strength between two electrons, that are
a distance $2d$ apart from each other, is just 1/8 of the one for
nearest-neighbor electrons. 
In the considered dipole limit, 
we achieve the same uniform interaction strength between neighboring
electrons and 
reduction for non-nearest-neighbor
$J$ couplings of Ref.~\cite{mchugh}, where, however, this result requires
a careful adjustment of both the inter-ion separation and 
end-trap strength.
\section{\label{gates} Universal quantum logic gates}
In this section we describe how to implement, in our system,
universal quantum computation.

As mentioned in the previous section, Hamiltonian (\ref{Heff}) is
substantially analogous to the one describing NMR molecules \cite{jones}. The
only difference is the fact that in the NMR systems we have
nuclear spins instead of electron spins. Therefore, universal quantum
processing with trapped electrons, in the presence of a magnetic
gradient, can be performed by using techniques similar to those
already developed in NMR experiments \cite{jones}.

In particular, Hamiltonian (\ref{Heff}) represents a set of $N$
electron spins, each one having a different precession frequency,
which interact by mutual couplings.
If we encode a qubit in the spin degree of freedom of each
particle, we can use electromagnetic pulses with appropriate
frequencies to selectively manipulate the information stored in
the spin state of each electron.
Hence, the detuning between the different electron frequencies
plays the same role of the chemical shift in NMR molecules.

Universal two-qubit gates, in our system, are achieved by means
of the mutual coupling between the electron spins. Indeed, this
interaction has the same form of the $J$-coupling between nuclear
spins in NMR molecules and can be used in a similar way to perform
two-qubit gates.

The set of unitary transformations consisting of single-qubit
gates plus C-NOT gates is computationally universal.
Let us describe, in detail, how to perform, in our system, these
operations.
If we apply a small transverse oscillating magnetic field resonant
with the spin precession frequency $\omega_{s0,j}$ of the $j$-th
electron
\begin{equation} \label{somf}
{\bf B_d} (t) = B_d[ {\bf{\hat{i}}} \cos(\omega_{s0,j} t + \theta)
                + {\bf{\hat{j}}} \sin(\omega_{s0,j} t + \theta) ],
\end{equation}
the relevant part of the system Hamiltonian becomes, in
interaction picture with respect to the unperturbed Hamiltonian
$H_{s0}=\sum_{i=1}^{N}(\hbar/2) \omega_{s0,i} \sigma_{z,i}$ and in
rotating wave approximation,
\begin{equation} \label{ship}
H_{IP}^{(spin)} \simeq \hbar \frac{\chi}{2} \left( \sigma_{+,j}
e^{-i \theta} + \sigma_{-,j} e^{i \theta} \right),
\end{equation}
where $\chi \equiv g\vert e \vert B_d / (2m_e)$ and $\sigma_{
\pm,j} \equiv ( \sigma_{x,j} \pm i\sigma_{y,j})/2$.\\
In the derivation of the above Hamiltonian we neglected the
spin-spin coupling terms present in Hamiltonian (\ref{Heff}), since
we assume $J_{i,j}\ll\chi$ for any $i,j=1,2,\ldots,N$.
Furthermore we suppose that the oscillating field is so small that
$B_d \ll b d_{i,j }$ for any $i,j=1,2,\ldots,N$. This last
condition, giving a Rabi frequency $\chi$ much smaller than the
detuning between the spin frequencies of the $i$-th and  $j$-th 
electrons, allows for the selective frequency addressing of each particle.

If the small transverse magnetic field is applied for a time $t$,
it produces a rotation on the spin state of the $j$-th particle
\begin{eqnarray} \label{nos1}
   \vert \downarrow \rangle_j &\rightarrow&
   \cos \left(\frac{\chi t}{2}\right)|\downarrow\rangle_j
    -i e^{-i \theta}\sin \left(\frac{\chi t}{2} \right) | \uparrow \rangle_j
, \\
   \label{nos2}
   | \uparrow \rangle_j &\rightarrow&
   \cos \left(\frac{\chi t}{2}\right) |\uparrow \rangle_j
   -i e^{i \theta} \sin \left(\frac{\chi t}{2}\right) |\downarrow \rangle_j
.
\end{eqnarray}
It can be shown that with an appropriate combination of these
operations, one can perform any single-qubit gate on the spin
qubit of the $j$-th electron. We define the interaction produced
by the Hamiltonian (\ref{ship}), applied for a time $t$, as a
$p_{s,j} (\chi t, \theta)$ pulse.
Hence, we can perform single-qubit gates on each electron, in a
selective way, by appropriately tuning the frequency of the
oscillating magnetic field.

Let us now consider the implementation of C-NOT gates.
A natural way to perform this two-qubit gate is to use a three
gate circuit. In particular, we apply in sequence, an Hadamard
gate on the target qubit, a controlled $\pi$ phase-shift gate
followed by 
another Hadamard gate on the target qubit.
However, it
is preferable to avoid Hadamard gates, as they are difficult to
implement. Hence, the Hadamard gates are conveniently replaced by an
inverse pseudo-Hadamard gate and a pseudo-Hadamard gate
\cite{jones}. These two gates are realized by applying
respectively a $p_s(\pi/2,\pi/2)$ pulse and a $p_s(\pi/2,-\pi/2)$
pulse.
The controlled $\pi$ phase-shift gate can be achieved, in
different ways, by applying appropriate sequences of $p_s$ pulses.
For example, in a two-qubit system, one possible sequence for
implementing this gate consists of four $p_s$ pulses and two
appropriately timed periods of free evolution \cite{jones}. In
particular, we apply: a free evolution of the system for a time of
$1/4J_{1,2}$, a $p_s(\pi,0)$ pulse, another free evolution of the
system for a time of $1/4J_{1,2}$, a $p_s(\pi/2,0)$ pulse, a
$p_s(\pi/2,\pi/2)$ pulse and a $p_s(\pi/2,0)$ pulse. All the above
pulses should be applied to both electron spins and the
application time of each $p_s$ pulse should be much smaller than
the time duration of each free evolution period.

With systems having more than two qubits, the efficient
implementation of C-NOT gates requires the application of more
complicated sequences of operations. This is necessary in order to
avoid errors due to the couplings with electrons not involved in
the gates.
Indeed, from Hamiltonian(\ref{Heff}) we see that there are mutual couplings
among all the particles of the system.
However, the sequence required to efficiently implement the C-NOT
gates, when we have more than two qubits, consists only of $p_s$
pulses acting on specific particles and appropriately timed
periods of free evolution \cite{jones2,leung}. In NMR context
this technique is known as refocusing.

We also remark that in our system the coupling strength decreases
rapidly with the increase of the inter-particle distance. Indeed
from equation (\ref{J}) we have $J_{i,j}\propto 1/d_{i,j}^3$. This
fact permits to simplify the refocusing sequences since the
interaction between distant electrons can be neglected
\cite{jones2,linden}. We recall that in this case the logic gates
between distant electrons can be performed by means of swap gates,
which move the quantum information among the particles. A swap
gate is realized by applying three C-NOT gates, where the two
qubits play alternatively the role of 
target and controller.

Let us give realistic estimates for the values of the electron
detunings and spin-spin couplings achievable in our system.
We consider a linear array of $10$ electrons with inter-particle
distance of the order of $1$mm. We suppose to apply a uniform
magnetic field of few Tesla, giving $\omega_s /2 \pi \approx 100$~GHz,
a magnetic gradient of about 50~T/m and assume an axial
frequency $\omega_z /2 \pi \approx 10$~MHz. In these conditions
we obtain a frequency detuning between neighboring particles of 
few MHz and a spin-spin coupling with strength $J \approx 20$~Hz.
These values for the detuning and the coupling strength are of the same
order of the corresponding quantities in NMR systems.
\section{\label{conclusion} Conclusion}
A novel concept of open planar Penning trap makes it possible to
design and build-up a scalable system for quantum computation,
consisting of trapped electrons in vacuum. Single particles are
confined in a linear array of traps, deposited on the same
substrate. A magnetic field gradient across the traps allows for
frequency addressing of each qubit, stored into the electron spin
as in NMR quantum computers. Moreover, the magnetic field gradient
couples internal (spin) and external (motional) degrees of freedom
of the same particle. Thanks to the Coulomb interaction among the
charged particles, this coupling effectively amounts to a direct
spin-spin interaction, with tunable coupling strength. 
In the limit of relatively large inter-electron spacing and
small axial oscillation amplitude, we obtain an analytical 
expression for the $J$ coupling, which allows for an immediate
evaluation of the interaction strength in terms of
the relevant system parameters (trap distance, axial frequency, 
and applied magnetic gradient). 
We emphasize that the $J$ coupling is proportional to $1/d^3$,
where $d$ is the inter-trap distance,
thus greatly reducing
the interaction between non-nearest-neighbor electrons.

This way,
the system of singly trapped electrons is formally identical to a
NMR molecule, suitable for quantum information processing. Hence,
the well established and developed techniques of NMR spectroscopy
can be extended and applied to our system. Qubit manipulation is
achieved via appropriate sequences of microwave pulses. The qubit
read-out is performed either by axial frequency detection, as in
conventional Pennning traps, or by capacity and charge
measurements as in semiconductor quantum dots.

The advantages over NMR systems are obvious: the number of qubits
is not limited neither by the molecule size -- the same substrate
can easily accomodate a large number of microtraps -- nor by the
frequency range -- the typical spin resonance frequency lies in the
GHz, whereas the detuning between neighboring particles is of the
order of a few kHz --, and the system is truly \emph{quantum} and,
at the same time, well
isolated from the environment. In addition, the geometry of the
system can be designed and optmized at will, using more
complicated two dimensional arrays, with possible applications to
the study and the simulation of quantum systems like the Ising
model.
Finally, Penning traps could be loaded with protons as well,
allowing the same NMR spectroscopy experiments with hydrogen,
but in a much more controlled and clean environment.
\begin{acknowledgments}
We acknowledge stimulating discussions on the practical 
implementation of this proposal with G. Werth and S. Stahl.
This research has been carried on in the frame of the project QUELE,
funded by the European Union under the contract
IST-FP6-003772.
\end{acknowledgments}

\begin{thebibliography}{99}
\bibitem{jones} J.A. Jones, Prog. NMR Spectroscop. \textbf{38},
         325 (2001).
\bibitem{leibfried1} D. Leibfried, R. Blatt, C. Monroe, and D. Wineland,
         Rev. Mod. Phys. \textbf{75}, 281 (2003).
\bibitem{rainerblatt} S. Gulde, M. Riebe, G.P.T. Lancaster, C. Becher,
         J. Eschner, H. H\"{a}ffner, F. Schmidt-Kaler, I.L. Chuang, and R.
         Blatt, Nature \textbf{421}, 48 (2003).
\bibitem{Leibfried} D. Leibfried, B. DeMarco, V. Meyer, D. Lucas, M.
         Barrett, J. Britton,
         W.M. Itano, B. Jelenkovic, C. Langer, T. Rosenband, and D.J.
         Wineland, Nature \textbf{422}, 412 (2003).
\bibitem{mintert} F. Mintert and Ch. Wunderlich, Phys. Rev. Lett.
         \textbf{87},
         257904 (2001).
\bibitem{wunderlich_1} Ch. Wunderlich, in \emph{Laser Physics at the Limit},
         p. 261 (Springer, Heidelberg, 2001).
\bibitem{wunderlich_2} Ch. Wunderlich and Ch. Balzer, in \emph{Advances in
         Atomic, Molecular and Optical Physics}, (Academic Press,
         San Diego, Ca, 2003).
\bibitem{wunderlich_3} Ch. Wunderlich, Ch. Balzer, T. Hannemann, F. Mintert,
         W. Neuhauser, D. Rei\ss, and P.E. Toschek, J. Phys. B:
         At. Mol. Opt. Phys. \textbf{36}, 1063 (2003).
\bibitem{mchugh} D. Mc Hugh and J. Twamley, Phys. Rev. A \textbf{71},
         012315 (2005).
\bibitem{ciaramicoli4} G. Ciaramicoli, I. Marzoli, and P. Tombesi,
         Phys. Rev. Lett. \textbf{91}, 017901 (2003).
\bibitem{ciaramicoli_5} G. Ciaramicoli, I. Marzoli, and P. Tombesi,
         Phys. Rev. A \textbf{70}, 032301 (2004).
\bibitem{stahl} S. Stahl, F. Galve, J. Alonso, S. Djekic, W. Quint,
         T. Valenzuela, J. Verd\`u, M. Vogel, and G. Werth,
         Eur. Phys. J. D \textbf{32}, 139 (2005).
\bibitem{bossBook} F. G. Major, V. Gheorghe, G. Werth, \emph{Charged
         Particle
         Traps}, (Springer, Heidelberg, 2005).
\bibitem{pritchard} E. A. Cornell, R. M. Weisskoff, K. R. Boyce
and D. E. Pritchard, Phys. Rev. A \textbf{41}, 312 (1990).
\bibitem{Verdu} J. Verd\'u et al., Physica Scripta \textbf{T112}, 68 (2004).
\bibitem{Djekic} S. Djekic et al., accepted for Eur. Phys. J. D.
\bibitem{patch} J.R. Anglin and W.H. Zurek, arXiv:quant-ph/9611049;
         C. Henkel, S. P\"{o}tting, and M. Wilkens,
         Appl. Phys. B \textbf{69}, 379 (1999).
\bibitem{diederich} M. Diederich et al., Hyperfine Interactions
        \textbf{115}, 185 (1998).
\bibitem{jackson} J.D. Jackson, \textit{Classical Electrodynamics}, 2nd
         ed., (Wiley, New York, 1975).
\bibitem{gabrielse} S. Peil and G. Gabrielse, Phys. Rev. Lett. \textbf{83}, 1287
        (1999).
\bibitem{brown} L.S. Brown and G. Gabrielse, Rev. Mod. Phys. \textbf{58},
         233 (1986).
\bibitem{powell} H.F. Powell, D.M. Segal, and R.C. Thompson, Phys.
        Rev. Lett. \textbf{89}, 093003 (2002).
\bibitem{jones2} J.A. Jones and E. Knill, J. Magn. Reson.
         \textbf{141}, 322 (1999).
\bibitem{leung} D.W. Leung, I.L. Chuang, F. Yamaguchi and Y.
         Yamamoto, Phys. Rev. A \textbf{61}, 042310 (2000).
\bibitem{linden} N. Linden, \={E}. Kup\u{c}e and R. Freeman, Chem.
         Phys. Lett. \textbf{311}, 321 (1999).
\end{thebibliography}

%
\end{document}